\documentclass[twocolumn,prd,amsmath,amssymb, aps,superscriptaddress, nofootinbib]{revtex4-1}
 \usepackage{comment}
\usepackage[titletoc,toc,title]{appendix}
\usepackage[version=4]{mhchem} 
\usepackage[normalem]{ulem}
\usepackage{mhchem}
\usepackage{amsmath}
\usepackage{amssymb}
\usepackage{amsfonts}
\usepackage{mathrsfs}
\usepackage{xcolor}
\usepackage{xfrac}
\usepackage{footmisc}
\usepackage{comment}
\usepackage{pifont}
\usepackage{times}
\usepackage[hidelinks]{hyperref}
\usepackage{enumitem}
\usepackage{centernot}
\usepackage{cancel}
\usepackage{slashed}
\usepackage{relsize}
\usepackage{times}
\usepackage{epsfig}
\usepackage{verbatim}
\usepackage{bm}
\usepackage[utf8]{inputenc}
\usepackage{graphics}
\usepackage{graphicx,color}
\usepackage{subfigure}
\usepackage[english]{babel}
\usepackage{adjustbox}
\usepackage{physics}
\usepackage{array}
\usepackage{orcidlink}

\definecolor{zima_blue}{HTML}{1393C1}
\hypersetup{setpagesize=false,bookmarksnumbered=true,
bookmarksopen=true,colorlinks=true,linkcolor=zima_blue,
urlcolor=zima_blue,citecolor=zima_blue,linktocpage=false}

\newcommand{\zb}[1]{{\color{black} {#1}}}

\newcommand{\dbd}[2]{\ifmmode \frac{\textrm{d}#1}{\textrm{d}#2}\else $\textrm{d}#1/\textrm{d}#2$\fi}
\newcommand{\pbp}[2]{\ifmmode \frac{\partial#1}{\partial#2}\else $\partial#1/\partial#2$\fi}

\DeclareMathAlphabet{\mathpzc}{OT1}{pzc}{m}{it}
 \newcommand{\eV}{\text{e\kern-0.15ex V}\xspace}

 \newcommand{\TeV}{\text{T\kern-0.1ex \eV}\xspace}

\DeclareMathAlphabet{\mathpzc}{OT1}{pzc}{m}{it}

\newcommand{\be}{\begin{equation}}
\newcommand{\ee}{\end{equation}}
\newcommand{\bea}{\begin{eqnarray}}
\newcommand{\eea}{\end{eqnarray}}

\begin{document}

\preprint{}

\title{Characterizing CJPL's Site-Specific Neutrino Floor as the Neutrino Fog Boundary
}

\author{Yingjie Fan}
\email{yingjiefan@s.ytu.edu.cn}
\affiliation{Department of Physics, Yantai University, Yantai 264005, China}

\author{Xuewen Liu~\orcidlink{0000-0003-3652-7237}}
\email{xuewenliu@ytu.edu.cn}
\thanks{correspongding author}
\affiliation{Department of Physics, Yantai University, Yantai 264005, China}

\author{Ning Zhou}
\email{nzhou@sjtu.edu.cn}
\thanks{correspongding author}
\affiliation{State Key Laboratory of Dark Matter Physics, Key Laboratory for Particle Astrophysics and Cosmology (MoE), Shanghai Key Laboratory for Particle Physics and Cosmology, Tsung-Dao Lee Institute \& School of Physics and Astronomy, Shanghai Jiao Tong University, Shanghai 201210, China}

\begin{abstract}
\zb{The neutrino floor, a theoretical sensitivity limit for dark matter (DM) direct detections}, 
is being redefined as the boundary of a dynamic ``neutrino fog", where neutrino signals become inevitable, obscuring DM detection due to the statistical and systematic uncertainties. 
This study provides the first site-specific analysis of the neutrino floor at China Jinping Underground Laboratory (CJPL), leveraging its unique geographic and environmental characteristics. 
We quantify how CJPL's suppressed atmospheric neutrino flux (around 30\% lower than \zb{Laboratori Nazionali del Gran Sasso (LNGS)}) 
reshapes the neutrino floor, thereby enabling improved sensitivity to high-mass WIMPs (mass $>10 \rm GeV$). Using a gradient-based framework, we derive CJPL's neutrino floor and estimate the detection prospects for the PandaX-xT experiment. Our results demonstrate that a 500 tonne-year exposure with PandaX-xT could touch the floor, probing spin independent cross-section down to $\sigma_{n}\sim 3\times 10^{-49} \rm cm^2$ \zb{at a DM mass of 70 GeV/$c^2$.}
\end{abstract}

\maketitle

\section{Introduction}
\label{sec:intro}
\zb{The search for dark matter (DM) has been one of the most significant endeavors in modern physics \cite{Cirelli:2024ssz}. Among the various DM candidates, Weakly Interacting Massive Particles (WIMPs) have been a leading hypothesis \cite{Bertone:2004pz,Goodman:1984dc}.} Direct detection experiments aim to observe the scattering of WIMPs off atomic nuclei in a detector. However, these experiments face a fundamental challenge in the form of the \textit{neutrino floor}. 

The neutrino floor, originally defined as the theoretical sensitivity limit for WIMPs in direct DM detection experiments, has long been regarded as an insurmountable barrier due to irreducible neutrino-induced backgrounds \cite{Billard:2013qya}. \zb{However, recent advances have revealed that there is no strict theoretical ``floor'' for DM direct searches, but a dynamic transition zone—termed the \textit{neutrino fog}—where statistical and systematic uncertainties obscure DM signals \cite{OHare:2021utq}.} 
This means that the neutrino fog delineates regions of parameter space where WIMP-nucleus interactions become indistinguishable from coherent elastic neutrino-nucleus scattering (CE$\nu$NS) events, primarily sourced from solar (e.g., $^8 \rm B$), atmospheric, and diffuse supernova neutrino fluxes.  
The new definition of neutrino fog emphasizes its statistical nature. The transition from Poisson-statistic-dominated to systematic uncertainty-limited regimes is quantified by the gradient index $n$, where $n=2$ marks the boundary of the fog (the neutrino floor). For $n>2$, sensitivity improvements require an exponential increase in exposure, making conventional detection strategies ineffective.

In fact, after numerous years of fruitless searches for DM particles, researchers have first detected a genuine signal emanating from a stream of neutrinos produced by Sun nuclear reactions. Notably, in 2024, the PandaX \cite{PandaX:2024muv} and XENON \cite{XENON:2024ijk} collaborations reported that their detectors have likely begun detecting this elusive neutrino fog \zb{($^8$B solar neutrinos)}.  
Recently, several theoretical studies on this subject have also been conducted, (as referenced in \cite{Tang:2023xub,Herrera:2023xun, Carew:2023qrj, Tang:2024prl, Bloch:2024suj, Blanco-Mas:2024ale, Maity:2024vkj}), \zb{ significantly advancing our understanding of the implications of this irreducible neutrino background for DM searches.} 

Critically, the morphology of the neutrino fog exhibits strong geographic dependence, as local neutrino flux variations—modulated by geomagnetic latitude, cosmic-ray modulation, and detector depth—directly influence background kinematics and systematics.

The China Jinping Underground Laboratory (CJPL), with its unique geographic profile ($18.06^\circ$N geomagnetic latitude, 2400 m rock overburden), presents an exceptional case for studying neutrino fog. 
Among all existing direct detection experiments, CJPL exhibits the largest crustal geoneutrino flux \cite{Wan:2016nhe} and the smallest reactor neutrino background \cite{Jinping:2016iiq}. More crucially, the atmospheric neutrino flux at CJPL is notably lower \cite{Zhuang:2021rsg}, which has a significant impact on the neutrino floor in the high-mass region of DM.
Considering these distinctive site-specific attributes, it is imperative to conduct localized fog calculations.

This study delves into two primary questions. Firstly, we examine how the unique neutrino flux characteristics at CJPL alter the landscape of the neutrino fog and floor in comparison to established benchmarks, such as xenon-based detectors located at the Laboratori Nazionali del Gran Sasso (LNGS). Secondly, we calculate the prospective sensitivities and estimate the necessary exposure to touch the neutrino floor in the PandaX experiment conducted at CJPL.

The structure of this paper is outlined as follows. In Section \ref{sec:flux}, we delve into the neutrino fluxes at CJPL. Section \ref{sec:NFs} is dedicated to deriving the neutrino floor and fog specific to CJPL. In Section \ref{sec:sensitivity}, we conduct calculation of the sensitivity of PandaX-xT experiment, aiming to determine the required exposure to reach the new neutrino floor in the high DM mass region. Finally, in Section \ref{sec:con}, we present our conclusions.

\section{NEUTRINO FLUXES at CJPL}
\label{sec:flux}

\begin{figure}[!htbp]
  \centering  \includegraphics[width=0.5\textwidth]{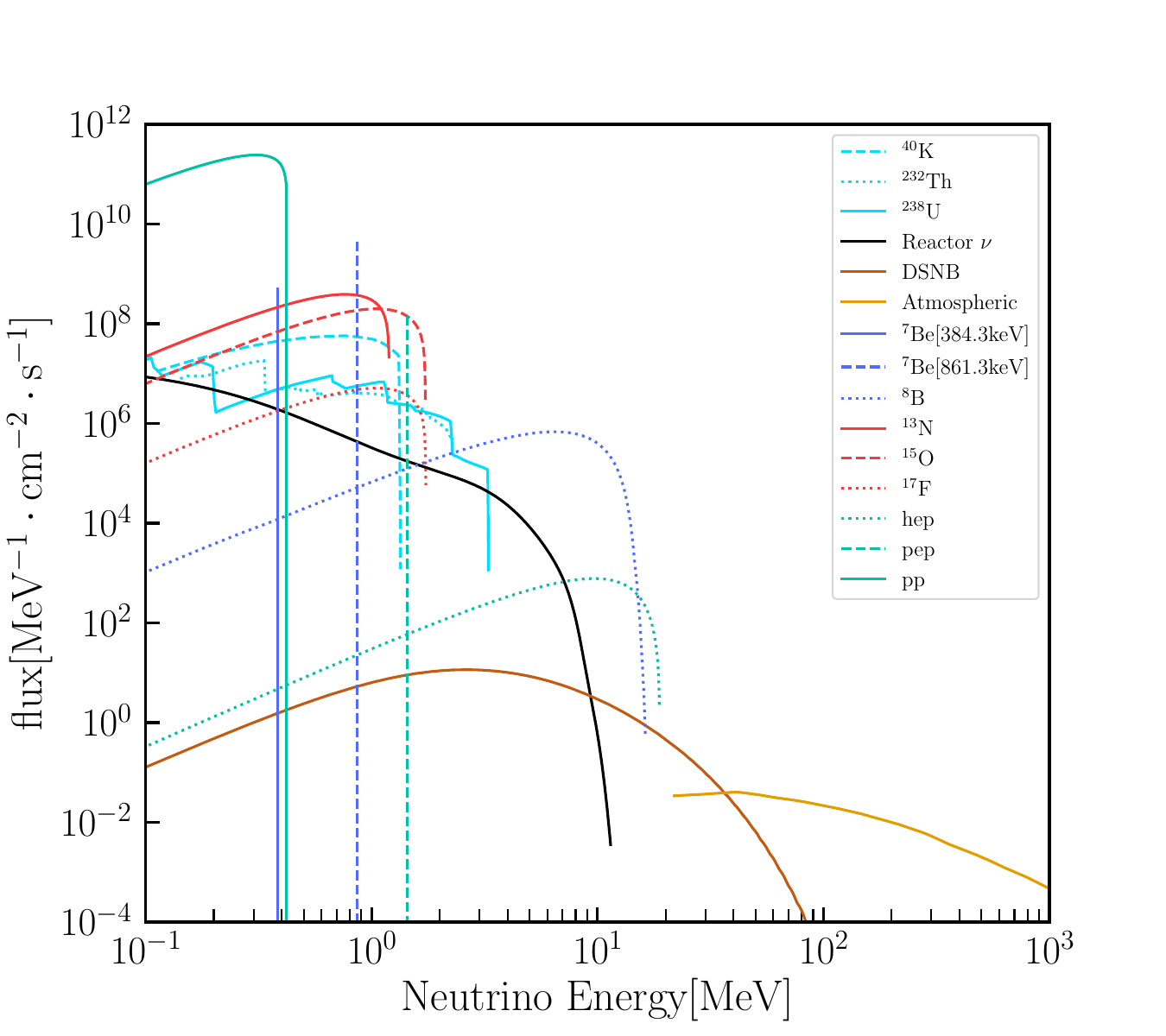}  
  \caption{
  \zb{Neutrino fluxes at CJPL. Fluxes from different neutrino sources are shown with different colors: geoneutrinos \cite{Gelmini:2018gqa} are light blue, neutrinos from reactors \cite{Jinping:2016iiq}  is shown with black, lines with purple are solar neutrinos \cite{Vinyoles:2016djt} , DSNB neutrinos \cite{OHare:2021utq}  are shown with brown, and atmospheric neutrinos \cite{Zhuang:2021rsg}  is shown with orange. }}
  \label{fig:nuflux}
\end{figure}

The neutrino fluxes, especially those of atmospheric neutrinos, geoneutrinos, and reactor neutrinos, are highly dependent on geographical locations. CJPL is uniquely positioned with several unparalleled features that distinguish it from other sites. It has the thickest overburden, providing exceptional shielding against cosmic rays, and has the lowest reactor neutrino flux due to its remote location from nuclear power plants \cite{IAEA}. Additionally, CJPL has the largest crustal geoneutrino flux, which is highly advantageous for geoneutrino studies. The laboratory also benefits from the lowest environmental radioactivity, ensuring a cleaner experimental environment, and has the longest solar neutrino path through the Earth, which is particularly beneficial for solar neutrino research. 
Therefore, the specific geographical and environmental features of CJPL contribute to its specialized neutrino background. 
We elaborate the pertinent constituents of the neutrino fluxes at CJPL, which are summarised in Fig. \ref{fig:nuflux}.

\begin{enumerate}
    
\item \textit{Solar neutrino}

The Sun produces neutrinos through two primary nuclear reaction chains: the proton-proton (pp) chain and the carbon-nitrogen-oxygen (CNO) cycle. Solar neutrinos dominate the flux at energies $E_\nu \leq 18.77$ MeV \cite{Adelberger:2010qa}. These neutrinos constitute the principal source of CE$\nu$NS events in DM detectors and constrain sensitivity to DM candidates with masses near $m_{\rm DM} \sim 10$ GeV. Very recently, the PandaX \cite{PandaX:2024muv} and XENON \cite{XENON:2024ijk} collaborations independently reported tentative observations of CE$\nu$NS signals.

In this study, we utilize the GS98 high-metallicity Standard Solar Model with the Barcelona 2016 calculations \cite{Vinyoles:2016djt}. For all components, we maintain the published normalization uncertainties, with the exception of  
$^{8}\text{B}$, for which a 2\% uncertainty was assigned based on comprehensive fits of global neutrino data \cite{Bergstrom:2016cbh}. Subsequent to  
$^{8}\text{B}$, the  
$^{7}\text{Be}$ electron-capture neutrino lines—which are pivotal for enhancing sensitivity to sub-GeV DM—are assigned a normalization uncertainty of 6\%.

\item \textit{Geoneutrinos}

Geoneutrinos are antielectron neutrinos  released during the decay processes of radioactive isotopes such as uranium (${}^{238}$U), thorium ( ${}^{232}$Th), and potassium (${}^{40}$K). Their flux directly reflects the radioactive heat production rate and thermal evolution history within the Earth's interior. 
The location-dependent geoneutrino flux is particularly sensitive to the amount of crust beneath the laboratory site, which contains the largest portion of heat producing elements. 
%
\zb{CJPL is situated near the Himalayan Mountains in China, where the crust is the thickest.} 
The uranium and thorium abundances in the local crustal composition significantly influence the spatial distribution of the geoneutrino flux in this region.
This has been previously identified as a favorable location for geoneutrino detection \cite{Wan:2016nhe}.

For concreteness, we use  geoneutrino flux from Ref. \cite{Gelmini:2018gqa}, with corresponding uncertainties for each component.

\item \textit{Reactor neutrinos}

This is another source of antineutrinos and influence the background at slightly higher masses. 
Jinping is far away from all the nuclear power plants \cite{IAEA} in operation and under construction.

The reactor neutrino background at Jinping is the lowest among all the direct detection experiments. 
The total differential reactor neutrino flux at Jinping is  employed from Ref. \cite{Jinping:2016iiq}.

\item \textit{The diffuse supernova neutrino background (DSNB)}

DSNB is a relic flux of neutrinos and antineutrinos from core-collapse supernovae in cosmic history—emerges as a critical astrophysical background for DM direct detection. 
\zb{For DM experiments, the DSNB-induced CE\(\nu\)NS creates an irreducible ``neutrino floor" near \(m_\chi \sim 20 \, \text{GeV}\) \cite{OHare:2021utq}.} A 50\% uncertainty on the all-flavor flux accounts for cosmic variance in supernova rates, progenitor mass-dependent spectral ambiguities, and neutrino oscillation effects \cite{Beacom:2010kk}. We use the same flux as in Ref. \cite{OHare:2021utq}.

\item \textit{Atmospheric neutrinos}

Atmospheric neutrinos exhibit significant geographic dependence due to variations in cosmic ray flux, Earth's geometry, and magnetic field effects. 
At high latitudes (e.g., polar regions), weaker geomagnetic shielding results in higher cosmic ray flux and neutrino production, while low latitudes (e.g., equatorial regions) experience reduced flux due to stronger magnetic deflection.

CJPL is situated in a geomagnetic low-latitude region (with a geomagnetic latitude of 18.06°N), where the atmospheric neutrino flux is predominantly driven by higher-energy cosmic rays. 
\zb{The flux at CJPL undergoes suppression in comparison to LNGS, primarily due to the relatively high rigidity cutoff energy. 
}

The atmospheric flux used in this work is the average of the solar min and solar max flux calculated in \cite{Zhuang:2021rsg}, with placing the recommended 25\% theoretical uncertainty.

\end{enumerate}

\section{Neutrino Floor and Fog for CJPL}
\label{sec:NFs}

In this section, we initially revisit the newly introduced definition of the neutrino floor, as presented in Ref. \cite{OHare:2021utq}. Subsequently, we derive our novel neutrino floor specifically for CJPL.

\zb{We focus on the standard spin-independent (SI) interactions between DM and nucleons and consider only elastic scattering. The scattering rate can be expressed as \cite{Schumann:2019eaa,DelNobile:2021wmp}:
\begin{eqnarray}\label{eq:DMrate}
\frac{\mathit{d} R}{\mathit{d} {E_R} } \left ( {E_R},t \right )
&=&{\sigma }_{n}\frac{\rho }{m_{\rm DM}} \frac{1}{2{\mu }_{n}^{2}} \sum_{T} {\zeta }_{T}{\left[Z+\left ( A-Z \right ){f}_{n}/{f}_{p}\right]}^{2}\notag\\
&\times&{F}_{SI}^{2}({E}_{R}){\eta}({v}_{min}\left({E}_{R}\right)), 
\end{eqnarray}
which is the scattering rate 
summed over all target nuclides. ${\sigma }_{n}$ is the WIMP-nucleon cross-section. In this work, we use $\rho=0.3~\rm GeV/cm^{3}$ for the local DM density, ${\mu }_{n}$ is the reduced mass of the WIMP-nucleon system, ${\zeta }_{T}$ is the mass fraction of the isotope $T$ in the detector, $Z$ is the atomic number of the target nucleus, $A$ is the atomic mass number of the target nucleus, and ${f}_{n}/{f}_{p}$ is the ratio between the neutron and proton couplings to the WIMP. For the standard assumption, we take ${f}_{p}={f}_{n}$, to ensure that the DM–proton and DM–neutron couplings are the isosinglet. ${F}_{SI}^{2}\left({E}_{R}\right)$ is taken to be the Helm form factor \cite{Helm:1956zz}. ${\eta}\left({v}_{min}\left({E}_{R}\right)\right)\equiv\int_{v\geq v_{min}} d\vec{v} f(\vec{v})/v$ is the integral of the DM velocity distribution.

For the sake of completeness, the rate for CE$\nu$NS can be found in Appendix \ref{app:eventrate}. Additionally, the case for the spin-dependent (SD) interaction is analyzed in Appendix \ref{app:SD}.
}

\subsection{Statistical Methods and Sensitivity Evolution}

The neutrino fog refers to the obscured region in direct DM detection experiments where the neutrino background and the DM signal energy spectra overlap significantly, rendering them indistinguishable. This phenomenon leads to a stagnation in experimental sensitivity. The underlying mechanism arises when the DM signal strength falls below the combined effects of systematic uncertainties and statistical fluctuations of the neutrino background, thereby masking the signal entirely. The neutrino floor, defined as the dynamical boundary of the neutrino fog, represents the critical hypersurface in parameter space where a DM signal transitions from a detectable regime to an indistinguishable one. This boundary is not static but evolves dynamically with experimental statistics, systematic errors, and target material properties.

The statistical inference employs a binned likelihood function, incorporating Poissonian probabilities and Gaussian distributions to model background interference:

\begin{eqnarray}
\mathscr{L}\left ( \sigma ,\Phi \right )= \displaystyle\prod_{i=1}^{{N}_{\mathrm{bins}}}\mathscr{P}\left [ {N}^{\rm obs}_{i}|{N}^{\chi }_{i}+
\displaystyle\sum_{j=1}^{{n}_{\nu }}{N}^{\nu }_{i}\left ( {\Phi }_{j}\right )\right ]\displaystyle\prod_{j=1}^{{n}_{\nu}}\mathscr{G}
\left ( {\Phi }_{j}\right ),
\end{eqnarray}
where \zb{
the Poissonian term describes the probability in each energy bin for the observed event count 
$N_{i}^{\text{obs}}$, given the expected signal events $N_{i}^{\chi}$ and the total expected neutrino events $N_{i}^{\nu}$ summed over neutrino fluxes	$\Phi_{j}$.} 
The Gaussian term introduces neutrino flux normalization parameters, which are considered as the nuisance parameters, with standard deviations $\delta\Phi$ quantifying systematic uncertainties in flux calculations. 
Parameters are the DM mass and cross section.

Hypothesis testing compares the null background-only model (\( \mathcal{M}_{\sigma=0} \)) against the signal-plus-background model (\( \mathcal{M} \)). A discovery threshold of \( q_{0} > 9 \) (corresponding to \( 3\sigma \) significance) defines the exclusion limit \cite{OHare:2021utq}, where the test statistic is defined as
\begin{eqnarray}\label{eq2}
{q}_{0}=\left\{\begin{matrix}-2\ln{\left [ \frac{\mathscr{L}\left ( 0,\hat{\hat{\mathbf{\Phi}}}|{\mathcal{M}}_{\sigma =0}\right )}{\mathscr{L}\left ( \hat{\sigma },\hat{{\mathbf{\Phi }}}|\mathcal{M}\right )}\right ]}
& \hat{\sigma}>0,\\ 
       0 & \hat{\sigma} \leq 0.
\end{matrix}\right.
\end{eqnarray}
\zb{We maximize the likelihood function at $\hat{\hat{\mathbf{\Phi}}}$ for the background-only model and at $( \hat{\sigma},\hat{\mathbf{\Phi}})$ for the signal-plus-background model. In this context, the Chernoff's theorem \cite{Chernoff:1954eli} applies, and under the null hypothesis where  $\mathcal{M}$ is true, the test statistic ${q}_{0}$ asymptotically follows a mixture distribution of $\frac{1}{2}{\chi }_{1}^{2} + \frac{1}{2}\delta\left( 0\right)$ 
 \cite{Algeri:2019lah}.
The statistical significance of the DM signal above background corresponds to 
$\sqrt{q_0}$.}

The sensitivity evolution as a function of $N$ (the number of observed background events) exhibits four distinct regimes\cite{OHare:2021utq}: (a) background independent regime ($N \ll1$). \zb{The sensitivity scales as $\sigma \propto N^{-1}$, where the neutrino background is small.} 
(b) Poisson statistic dominated regime ($N\gg1$). The sensitivity follows $ \sigma \propto N^{-1/2}$, governed by statistical fluctuations.  
(c) systematics uncertainty dominated regime ($N \geq 1/\delta\Phi^{2}$). The sensitivity stagnates as $\sigma\propto \sqrt{{(1 + N\delta\Phi^{2})}/{N}}$, where systematic errors dominate \cite{Billard:2013qya}. 
\zb{(d) When exposures are very high and observed event counts large, the experiment can effectively measure its own background, reinstating the Poissonian regime.
}

With these features, resulting the mathematical characterization of the neutrino fog, known as the `opacity', which is defined by the gradient index:

\begin{eqnarray}
n=-\big({\frac{\mathit{d}\log{\sigma}}{\mathit{d}\log{N}}}\big)^{-1}.
\end{eqnarray}
For $n=2$ (Poisson statistic dominated regime), the sensitivity adheres to $\sigma \propto N^{-1/2}$.  
When $n > 2$ (systematics uncertainty dominated regime), the experiment enters the neutrino fog’s saturation phase, where sensitivity plateaus.  

By mapping the $n = 2$ contour, the neutrino floor is identified as the critical hypersurface in the DM mass $m_{\chi}$ and cross-section $\sigma$ parameter space where sensitivity degradation becomes irreversible.

\subsection{Neutrino floor for CJPL}

\begin{figure}[!htbp]
\centering
\includegraphics[width=0.5\textwidth]{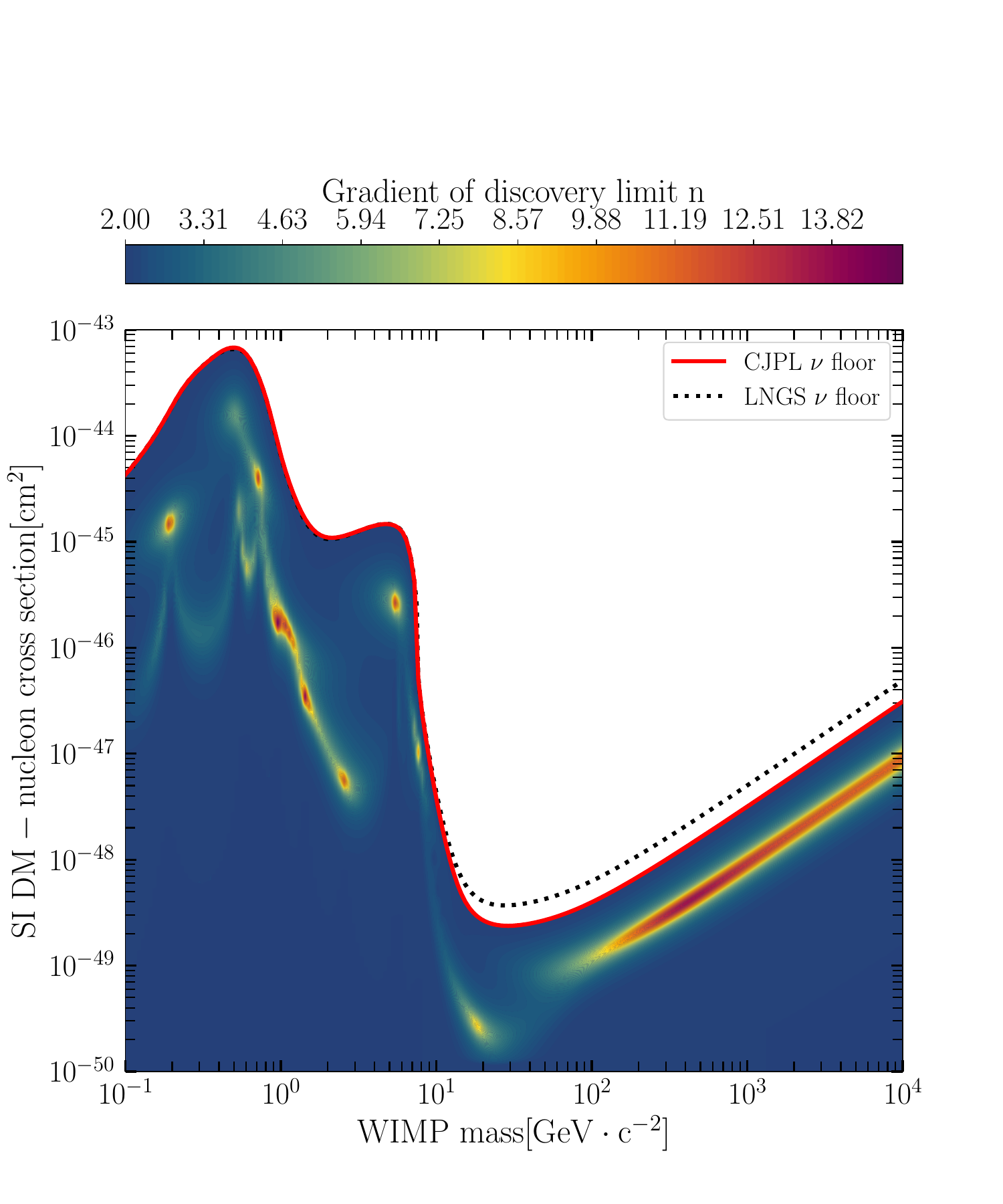}
\caption{Neutrino floor and fog for CJPL.  The red dashed curve represents the neutrino floor as the boundary of the neutrino fog with $n=2$ in this work, the black dotted line is the neutrino floor at LNGS 
derived in Ref. \cite{OHare:2021utq}. 
The color scale shows the value of $n$.} 
\label{fig:3floors}
\end{figure}

Building upon the public code \cite{NeutrinoFog-code} developed by O'Hare, this study pioneers the integration of neutrino flux data from CJPL to derive geographically characterized neutrino floor curves, as shown in Fig.~\ref{fig:3floors}. 
The comparative analysis reveals key findings through following comparisons.

For the location effects, in the high-mass region (WIMP mass $>10$ GeV), CJPL's neutrino floor demonstrates a $30$\% lower background limit compared to LNGS, marked by the red solid curve. \zb{For LNGS, the neutrino fluxes and uncertainties used are identical to those in the calculation of Ref.~\cite{OHare:2021utq}.} This advantage stems from CJPL's unique geological profile: Its atmospheric neutrino flux is reduced by approximately {($20-40$)\% relative to LNGS}, attributable to its lower geomagnetic latitude. Consequently, high-energy nuclear recoil backgrounds—dominant in this mass range—are suppressed. 
Future research should aim to optimize the treatment of atmospheric neutrinos to improve the accuracy of neutrino background limits.

In the realm below 10 GeV, the neutrino floors at CJPL and LNGS display a striking degree of consistency. This convergence is primarily attributed to the dominance of solar neutrinos, which contribute an impressive 90\% to the overall neutrino flux, with minimal variations observed across different geographical locations. 
Although the crustal geoneutrino flux at CJPL is notably higher than that at LNGS—a disparity linked to regional variations in uranium and thorium concentrations—
the impact of geoneutrinos in the low-mass region is rendered negligible.

To visualize the neutrino fog,  the color-mapping technique across the DM parameter space based on the gradient index \( n \) is implemented. This methodology is demonstrated in Fig. \ref{fig:3floors}, where the color scale above the plot explicitly indicates the \( n \)-value for each point within the fog.  

The opacity of the neutrino fog quantifies the resistance to experimental progress through the parameter space, revealing regions where overlapping neutrino backgrounds obscure DM discovery. This metric highlights zones where spectral degeneracies between DM and neutrino-induced recoils are most pronounced. Darker regions in Fig. \ref{fig:3floors} correspond to  $n > 2$, signifying enhanced spectral degeneracy where DM and neutrino event rates become nearly indistinguishable.  

While subtle differences between DM and neutrino signals persist in most scenarios, these distinctions can be statistically resolved in high-exposure regimes. Notably, once the accumulated event count \( N \) surpasses a critical threshold, the sensitivity scaling reverts to \( n = 2 \) for extremely low cross-sections, reflecting a transition back to Poisson-statistics-dominated sensitivity.  

A comparison reveals distinct characteristics between the neutrino fog at CJPL and LNGS. These differences stem from site-specific variations in neutrino flux components and their associated systematic uncertainties. Such geographic disparities underscore the necessity of site-specific fog modeling to optimize future DM detection strategies.  

\section{EXPERIMENTAL SENSITIVITY of PandaX-xT}
\label{sec:sensitivity}

Given the establishment of the new neutrino floor at CJPL, it is crucial to explore the potential detection prospects by the PandaX experiment. 
We undertake an estimate of the direct detection capabilities using the PandaX-xT experiment at CJPL \cite{PANDA-X:2024dlo}, focusing on a liquid xenon detector with a mass of 40 tonnes. Our calculation assumes a duration of exposure spanning several years, as detailed in \cite{PANDA-X:2024dlo}.

To properly reproduce the event rate measured by the experiments, we need to take into account the detection efficiency, the total number of events is given by
\begin{equation}
  N\approx exposure\times\int_{0}^{\infty}\mathit{d}{E}_{R}
  \frac{\mathit{d} R}{\mathit{d} {E_R} } \epsilon \left ( {E_R} \right ). 
\end{equation}
Here,  $\epsilon \left ( {E_R} \right )$is the detection efficiency, we use the efficiency curve of the current PandaX-4T detector~\cite{PandaX-4T:2021bab}, which covers the energy range from approximately 4 to 110 keV. 
To ensure the completeness of all detectable events, we set the upper limit of the energy window at a sufficiently large value, covering the maximum energy range that the experiment can detect.

Based on the aforementioned statistical methods, the likelihood function is modified as follows to perform the detection simulation:

\begin{eqnarray}
\mathscr{L}\left ( \sigma ,\Phi \right )&= \displaystyle\prod_{i=1}^{{N}_{\mathrm{bins}}}\mathscr{P}\left [ {N}^{\rm obs}_{i}|{N}^{\chi }_{i}+
\displaystyle\sum_{j=1}^{{n}_{\nu }}{n}_{\nu }^{i}\left ( {\Phi }^{j}\right )+{n}_{i}^{\rm bkg}\right ]\notag\\
&\times\displaystyle\prod_{j=1}^{{n}_{\nu}}\mathscr{G}\left ( {\Phi }_{j}\right )\mathscr{G}\left ( {n}_{i}^{\rm bkg}\right )
\end{eqnarray}
where ${n}_{i}^{\rm bkg}$ represents the background events in the PandaX detector, following the results reported in \cite{PANDA-X:2024dlo}. Additionally, we assume that the background events are uniformly distributed within the detector. It should be noted that we have updated the neutrino nuclear recoil background event rate using our calculated results for the front part of this work. 

\begin{figure}[!htbp]
  \centering
  \includegraphics[width=0.5\textwidth]{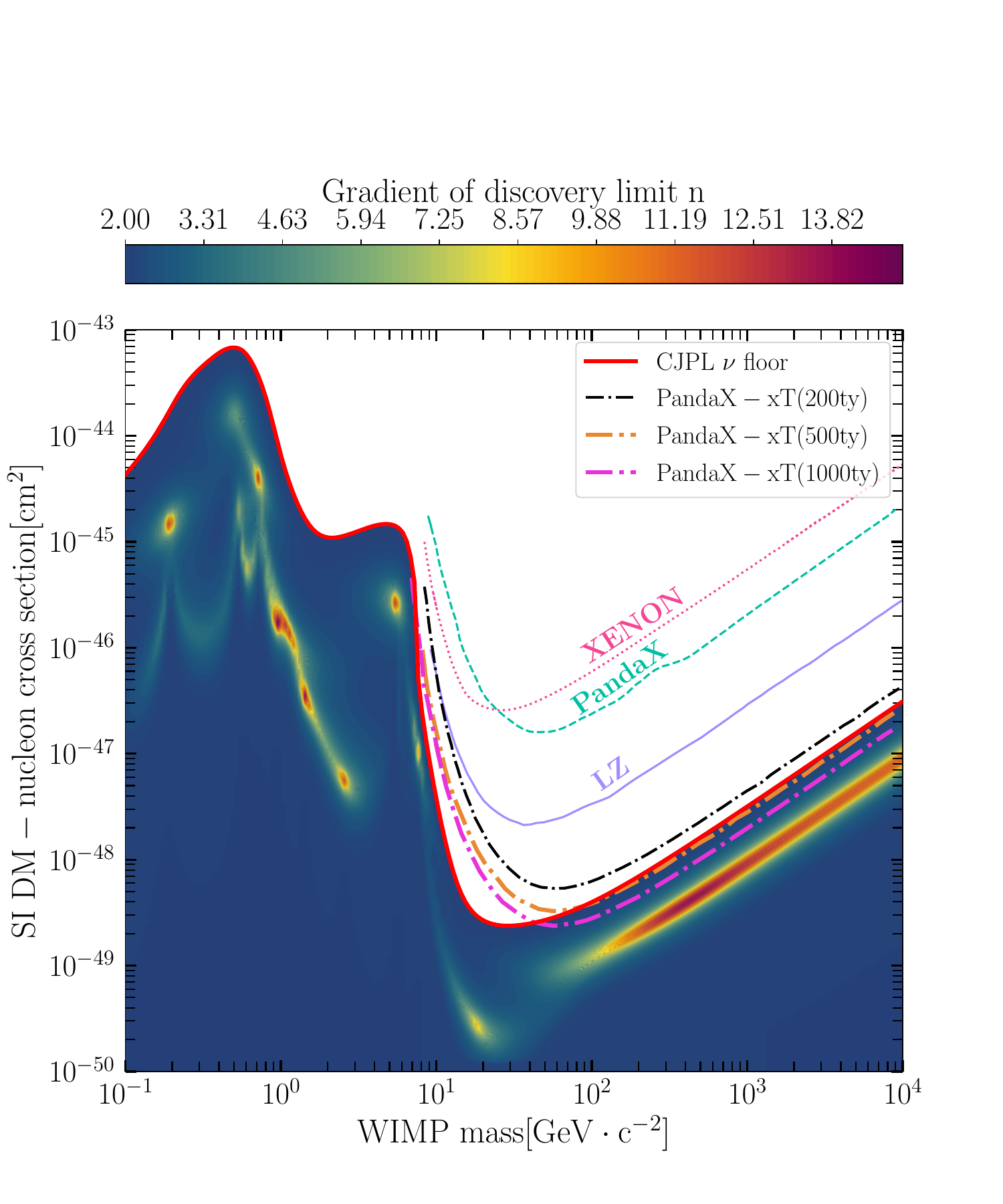} 
  \caption{Sensitivity and neutrino background. The latest excluded limits are shown with purple solid line for LZ \cite{LZ:2024zvo}, PandaX4T \cite{PandaX:2024qfu} with green dashed line and XENONnT \cite{XENON:2023cxc} with magenta dotted line. The prospective sensitivities are also shown in this figure, the black dash-dotted line is the prospect from PandaX-xT with a 200 ty exposure which is consisted with Ref. \cite{PANDA-X:2024dlo}, orange dash dotted line and pink solid line corresponding to the sensitivity for a 500 ty and 1000 ty exposure at PandaX-xT experiment respectively.}
  \label{fig:sensitivity}
\end{figure}

Applying the likelihood ratio test (Eq. \ref{eq2}), 
the projected 90\% confidence level exclusion sensitivity reaches of the PandaX-xT  experiment  are shown in Figure~\ref{fig:sensitivity}. 
For benchmark validation, we initially compute the sensitivity curve at 200 tonne-years (ty) exposure (black dot-dashed line), which shows well consistency with the PandaX Collaboration's baseline simulations \cite{PANDA-X:2024dlo}. This agreement confirms the validity of our treatment.
The the orange and pink dot-dashed curves are the sensitivities with 500 ty, and 1000 ty exposures respectively. 
\zb{We also include the neutrino floor and fog in the plot. When the exclusion sensitivity touches the neutrino floor, further improvement would need a significant increase of the experimental exposure.}

\zb{For WIMP mass smaller than 10 GeV/\(c^{2}\), the sensitivity has almost no improvement with more exposure, this is due to the much lower detection efficiency in this region. In future detections, the PandaX-xT experiment would achieve a much better sensitivity in this low mass region by lowering the detection threshold~\cite{PandaX:2024muv}. Specially, it is expected to measure the solar \(^{8}\)B solar neutrinos with a precision better than 10\%~\cite{PANDA-X:2024dlo}.}

Our central finding reveals that penetrating the neutrino floor requires accumulated exposures exceeding 500 ty for WIMP masses \zb{above 70 GeV/\(c^{2}\) (see the orange dot-dashed curve)}. At this threshold, PandaX-xT's sensitivity approaches cross-sections of \(\sigma_{n} \sim 3\times 10^{-49}\)~cm\(^2\) at a WIMP mass of 70 GeV/\(c^{2}\), marking the onset of neutrino floor dominance.
We further present the results for a 1000 ty exposure. In the ``saturation region" of the neutrino fog, systematic uncertainties—such as those associated with the calculation of the neutrino flux—become the dominant factors. Consequently, the experimental sensitivity cannot be effectively improved with increasing exposure.

\section{Conclusion}
\label{sec:con}

This study presents the first site-specific characterization of the neutrino floor and fog at CJPL, driven by its unique geographic and environmental conditions. By integrating the suppressed atmospheric neutrino flux of CJPL($\sim30\%$ lower than LNGS) and the enhanced geoneutrino contributions into a gradient-based statistical framework, we redefine the neutrino fog boundary as a dynamic transition zone governed by Poisson-statistical and systematic uncertainties.
These features reduce the neutrino floor by nearly 30\% for high-mass WIMPs (\(m_{\rm DM} > 10 \, \text{GeV}\)), enhancing sensitivity through reduced neutrino backgrounds. Below \(10 \, \text{GeV}\), solar neutrino dominance homogenizes sensitivity limits across sites.  
The sensitivity estimation of the PandaX-xT experiment shows that a 500 ty exposure reaches cross-sections as low as \(\sigma_{n} \sim 3\times 10^{-49} \, \text{cm}^2\) at a WIMP mass of 70 GeV$/c^2$, intersecting CJPL’s neutrino floor. Critical challenges persist, particularly systematic uncertainties in neutrino flux normalization, underscoring the need for refined models and multi-detector synergies.

\section*{Acknowledgments}

The work by X.L. is supported by the Project of Shandong Province Higher Educational Science and Technology Program under Grants No. 2022KJ271. The work by N.Z. is supported by National Science Foundation of China (No. 12325505).

\appendix


\section{Scattering rate for CE$\nu$NS}
\label{app:eventrate}

Neutrinos can scatter elastically off nuclei and produce recoils with very similar spectra to the ones found by DM-nucleus scattering. 
\zb{Currently, measurements of CE$\nu$NS have been performed by the COHERENT collaboration~\cite{COHERENT:2017ipa,COHERENT:2020iec} and
 the CONUS+ experiment \cite{Ackermann:2025obx},} and it is also well-understood in the Standard Model~\cite{Freedman:1973yd,Freedman:1977xn}. Similar to the WIMP event rate calculation, the neutrino event rate is computed by the convolution of the differential CE$\nu$NS cross section and the neutrino flux,
\begin{eqnarray}
    \frac{\mathit{d}{R}_{\nu }}{\mathit{d}{E}_{r}}= \frac{1}{{m}_{N}}\int_{{E}_{\nu}^{min}}\frac{\mathit{d}\Phi}{\mathit{d}{E}_{\nu}}\frac{\mathit{d}
    {\sigma}_{\nu N}\left({E}_{\nu}\right)}{\mathit{d}{E}_{r}}\mathit{d}{E}_{r}, 
\end{eqnarray} 
here we cut off the integral at the minimum neutrino energy that can cause a recoil with $E_r$: $E_{\nu}^{\min }=\sqrt{m_{N} E_{r} / 2}$.
The differential neutrino-nucleus cross section as a function of the recoil energy and the neutrino energy is given by~Refs.~\cite{Scholberg:2005qs,Billard:2013qya,Ruppin:2014bra,OHare:2020lva},
\begin{equation}
\frac{d\sigma(E_{\nu},E_r)}{dE_r} = \frac{G_f^2}{4\pi}Q_{\omega}^2m_N\left(1-\frac{m_NE_r}{2E_{\nu}^2}\right)F^2_{SI}(E_r), 
\end{equation}
where $m_N$ is the nucleus mass, $G_f$ is the Fermi coupling constant and $Q_{\omega} = N - (1-4\sin^2\theta_{\omega})Z$ is the weak nuclear hypercharge with $N$ the number of neutrons, $Z$ the number of protons, and $\theta_{\omega}$ the weak mixing angle, taking $\sin^2{\theta_W} = 0.2387$. The presence of the form factors describes the loss of coherence at higher momentum transfer and is assumed to be the same as for the WIMP-nucleus SI scattering, for which we use the standard Helm form factor~\cite{Lewin:1995rx}.

\section{Neutrino fog and sensitivity of PandaX-xT for SD interaction at CJPL}
\label{app:SD}

\zb{
In general, the cross section for SD interaction between DM and nucleus can be written in the following form \cite{Schumann:2019eaa,Yang:2013txa}:
\begin{eqnarray}
\sigma _{N }^{SD} & = & \frac{32}{\pi} G_{F}^{2}{\mu _{N}^{2}} \left ( a_{p} \left \langle S_{p}  \right \rangle  + a_{n}\left \langle S_{n}  \right \rangle  \right ) ^{2} \frac{J+1}{J}, 
\end{eqnarray}
where $J$ is the total nuclear angular momentum, $G_{F}$ is the Fermi coupling constant,  $\mu_{N}$ is the reduced mass of the DM-nucleus system. $\left \langle S_{p}  \right \rangle$ and $\left \langle S_{n}  \right \rangle$ are the expectation values of the total spin operators for proton and neutron \cite{Menendez:2012tm}. $a_{p,n}$ are the couplings to proton and neutron. In typical experimental setups, the scattering cross-section of the entire nucleus with DM is considered to involve only protons or neutrons within the nucleus, by setting $a_{p}$ or $a_{n}$ to zero respectively. Then we have 
\begin{eqnarray}
\sigma _{N (p,n) }^{SD} & = & \frac{4}{3} \frac{\mu _{N}^{2}}{\mu _{p,n}^{2}} \left \langle S_{p,n}  \right \rangle  ^{2} \frac{J+1}{J}\sigma _{p,n}, 
\end{eqnarray}
with $\sigma _{p}$ and $\sigma _{n}$ are scattering cross sections of protons or neutrons with DM.}
 \begin{figure*}[!htbp]
  \centering
  \subfigure[]{
  \includegraphics[width=0.485\textwidth]{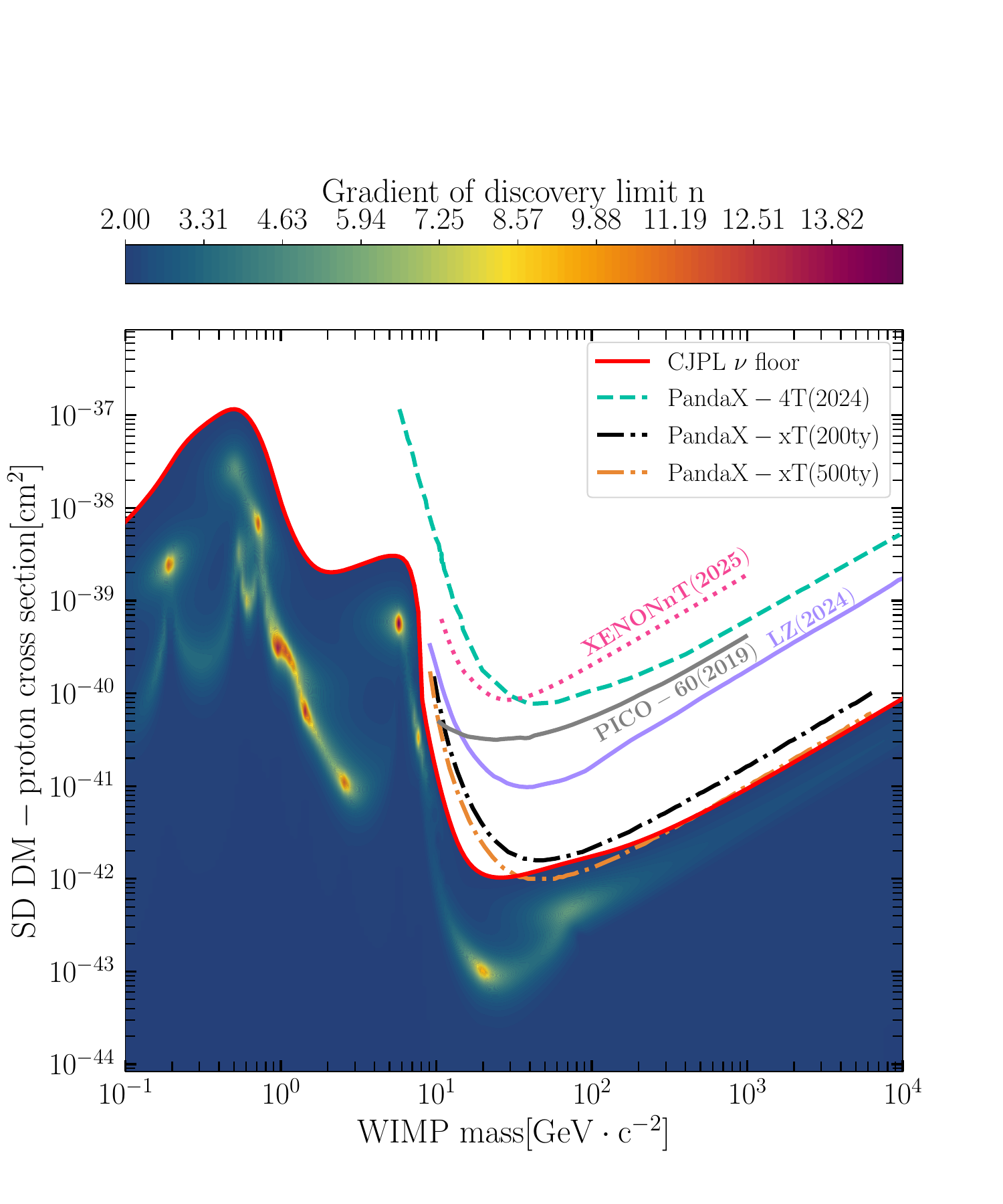}}
\subfigure[]{
  \includegraphics[width=0.485\textwidth]{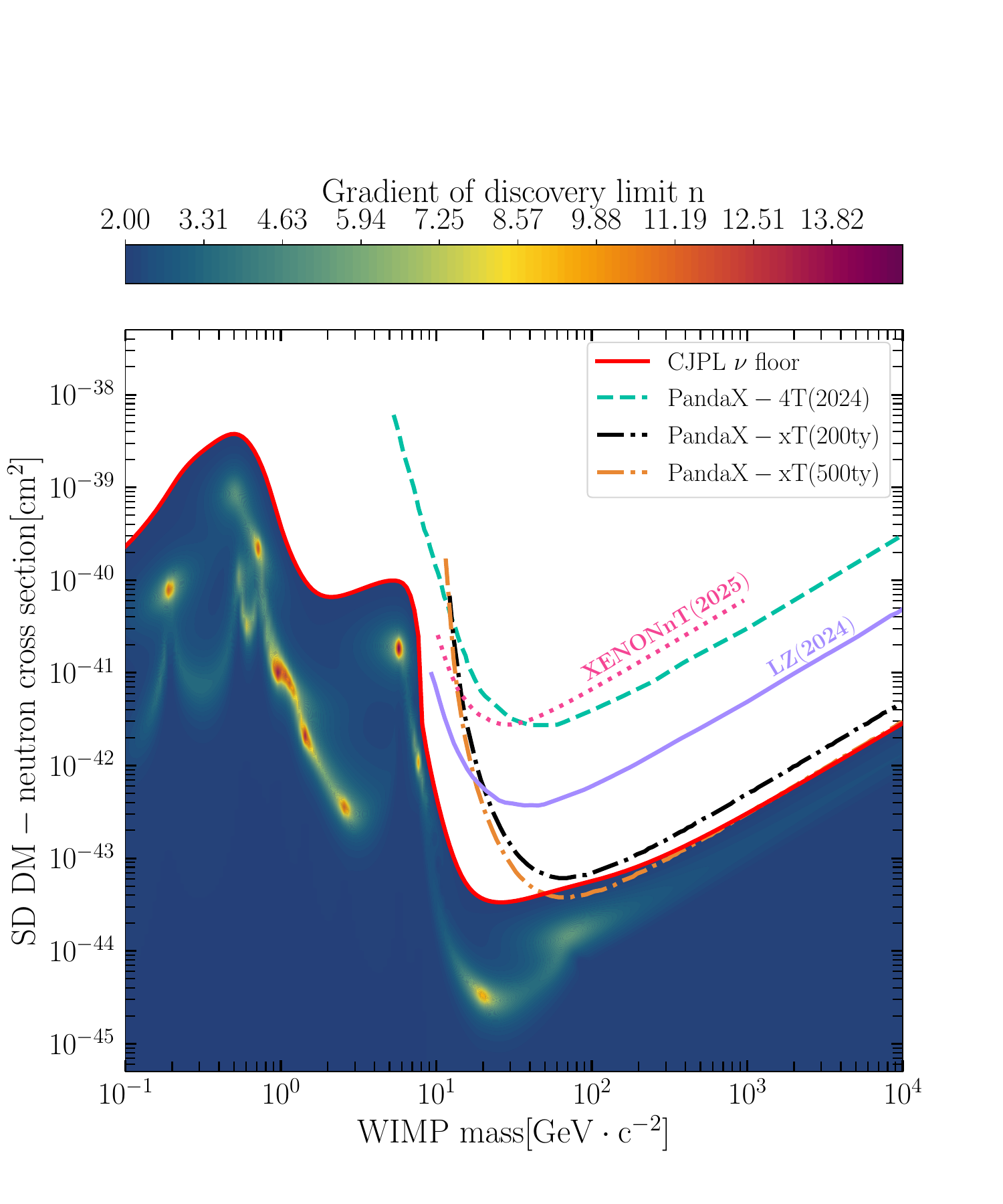}}
  \caption{\zb{Neutrino fog and exclusion sensitivities for SD DM-nucleon interactions at CJPL: (a) DM-proton interaction; (b) DM-neutron interaction. The latest exclusion limits are shown in each panel: purple solid line for LZ \cite{LZ:2024zvo}, a green dashed line for PandaX-4T \cite{PandaX:2024qfu}, a grey solid line for PICO-60 \cite{PICO:2019vsc}, and a pink dotted line for XENONnT \cite{XENON:2025vwd}. Additionally, the black (orange) dot-dashed lines in both (a) and (b) show the sensitivity curves of PandaX-xT for a 200(500) ty exposure.}}
  \label{fig:NuFog_SD}
\end{figure*}

\zb{
Employing the same procedure used in the SI case above, we derive the neutrino fog and the sensitivities of the PandaX-xT experiment for DM-proton/neutron SD interactions; these are presented in Figure \ref{fig:NuFog_SD}. Additionally, the latest excluded limits are shown with a purple solid line for LZ \cite{LZ:2024zvo}, a green dashed line for PandaX-4T \cite{PandaX:2024qfu}, a grey solid line for PICO-60 \cite{PICO:2019vsc}, and a pink dotted line for XENONnT \cite{XENON:2025vwd}. 

Our results reveal that  PandaX-xT detector with a 500 ty exposure  will encounter neutrino fog for both the DM-proton and DM-neutron SD scattering, presented by the orange dot-dashed lines in Figure \ref{fig:NuFog_SD}. 

}

\bibliography{biblio}
\end{document}